\documentclass[epj]{svjour}
\usepackage{latexsym}
\usepackage{graphicx}
\usepackage[latin1]{inputenc}
\usepackage{amsmath}
\begin{document}
\title{Anomalous photon diffusion in atomic vapors}

\author{M. Chevrollier\inst{1}\thanks{martine@otica.ufpb.br} \and N. Mercadier\inst{2} \and W. Guerin\inst{2}\thanks{Present address: Physikalisches Institut, Eberhard-Karls-Universit\"{a}t T\"{u}bingen, Auf der Morgenstelle 14, D-72076 T\"{u}bingen, Germany} \and R. Kaiser\inst{2}}

\institute{Laborat\'{o}rio de Espectroscopia \'{O}tica, Departamento de F\'{i}sica, Universidade Federal da Para\'{i}ba, Cx. Postal 5008 \\ 58051-900 Jo\~{a}o Pessoa, PB, Brazil \and Institut Non Lin\'{e}aire de Nice, CNRS and
Universit\'e Nice Sophia-Antipolis, 1361 route des Lucioles, 06560
Valbonne, France}
\date{This draft: \today / Received:  / Revised version: }
%
\abstract{The multiple scattering of photons in a hot, resonant, atomic vapor is investigated and shown to exhibit a L\'{e}vy Flight-like behavior. Monte Carlo simulations give insights into the frequency redistribution process that originates the long steps characteristic of this class of random walk phenomena.
} 
\maketitle
\section{Introduction}
\label{intro}
Random walk processes of particles, such as the Brownian motion, are usually described as Gaussian stochastic processes, where the distribution $P(x)$ of the size $x$ of the random steps has finite first and second moment, so that, according to the Central Limit Theorem (in the limit of a large number of steps), the total displacement, $r$, distribution converges towards a Gaussian and the process is said to be normal or diffusive. Properties such as the square root dependence of the total displacement with time characterize this so-called normal diffusion ($\gamma =1$ in Eq. \ref{displacement}):
\begin{equation}
<r^2>=D t^{\gamma}.
\label{displacement}
\end{equation}

The 1990's saw the emergence of a new class of systems which do not follow the Central Limit Theorem, and whose dynamics is dominated by large but rare random events. Systems exhibiting long-tailed step distributions with diverging second moment give rise to superdiffusion, with $\gamma > 1$ in Eq. \ref{displacement}. If this step size distribution asymptotically scales like a power-law $P(x)\propto x^{-\alpha}$ with $\alpha < 3$, the total displacement profile converges towards a L\'{e}vy distribution after a large number of steps \cite{Levy1937}. Note that, if $\alpha < 2$, even the first moment of the step size distribution diverges, i.e. the mean-free-path between two scattering events cannot be defined.

The number of physical phenomena shown to follow L\'{e}vy instead of Gaussian statistics has been steadily growing in the last 20 years \cite{Bouchaud1990,Shlesinger1995,Metzler2000}. Recently,  L\'{e}vy processes involving light propagation have been demonstrated in an optical material engineered to induce photon superdiffusive behavior \cite{Wiersma2008}. In such a system, long steps take their origin in inhomogeneities of spatial order artificially introduced in the material. As we shall show, another, much simpler, photonic system that exhibits Lévy flights, is a hot atomic vapor. In that case however, spatial L\'{e}vy flights of resonant light take their origin in spectral inhomogeneities.

In this paper, we present our study of Lévy flights of photons in atomic vapors. Contrary to most previous work, we measure directly the microscopic ingredient of the superdiffusive behavior, i.e. the single step-size distribution. In the next section, we present the physical origin of the Lévy flights as well as our experimental methods and measurements. Then, in section \ref{Monte-Carlo}, we detail our numerical simulations of such a system.

\section{Photon diffusion in atomic vapors}
\label{sec:AA}

\subsection{Origin of long steps}

Radiation trapping in an atomic vapor is a well known phenomenon \cite{Kenty1932,Holstein1947,Molisch1998}, whereby atoms in a vapor absorb resonant radiation from an incident source and re-emit photons that can be re-absorbed in the vapor, and so on. This photon multiple scattering happens in atomic media ranging from laboratory atomic vapors in optical cells to interstellar space. Under simple hypotheses, one can show that the distribution for the length covered by a photon between an emission and an absorption (jump size distribution) asymptotically decays like a power law of diverging second moment, characterizing resonant photons trajectories in atomic vapors as L\'{e}vy flights \cite{Pereira2004}. The diffusion of photons exhibits an anomalous behavior through the inhomogeneity introduced in the scattering process by the frequency dependence of the absorption coefficient and the frequency redistribution due to the atoms velocity.

In a dilute atomic vapor of two level atoms where the Doppler broadening of the atomic lineshape is negligible in comparison with the natural width, as with laser-cooled atomic vapors \cite{Fioretti1998,Labeyrie2003,Labeyrie2005,Pierrat2009}, the scattering is quasi elastic, i.e. the frequency of the re-emitted photon is almost the same as the frequency of the absorbed one, so that frequency redistribution can be neglected. For light at a frequency $\nu$, the jump size distribution then writes
\begin{equation}
P(x,\nu)=-\frac{\partial T(x,\nu)}{\partial x}= k_a(\nu) e^{-k_a(\nu) x},
\label{PdeX}
\end{equation}
where $k_a(\nu)$ is the absorption profile and $T(x,\nu)$ the probability of a photon of frequency $\nu$ going through a distance $x$ without being absorbed,
\begin{equation}
T(x,\nu)=e^{-k_a(\nu)x}.
\label{Beer}
\end{equation}
In the general case where inelastic scattering cannot be discarded, as in normal laboratory temperature condition, however, the jump size distribution function has to be computed by averaging over the spectrum of emitted light $k_e$:
\begin{equation}
P(x)= \int_{0}^{+\infty} k_e(\nu) k_a(\nu) e^{-k_a(\nu) x} d\nu \; .
\label{PdeXaverage}
\end{equation}

In presence of Doppler broadening alone, the absorption profile is gaussian
\begin{equation}
k_D(\nu)= N \sigma_0 \frac{\Gamma}{4}\frac{c}{\sqrt{\pi} v_0 \nu_0} e^{-\chi(\nu)^2} \; ,
\label{Doppler}
\end{equation}
while it is lorentzian for natural broadening
\begin{equation}
k_L(\nu)= N \sigma_0 \frac{1}{1+\left[ 4\pi (\nu -\nu_{0})/\Gamma \right] ^{2}} \; , \label{Lorentz}
\end{equation}
and Voigt if the two mechanisms are mixed,
\begin{equation}
k_V(\nu)= N \sigma_0 \frac{\Gamma}{4}\frac{a}{\pi}\frac{c}{\sqrt{\pi} v_0 \nu_0} \int_{-\infty }^{+\infty }\frac{\exp (-y^{2})}{a^{2}+(\chi(\nu)-y)^{2}}dy \; .
\end{equation}
Here, $N$ is the atomic density, $\sigma_0$ the on-resonance atomic cross section for atoms at rest, $\nu_0$ the resonance frequency, and $\Gamma$ the natural width of the atomic transition. The most probable velocity at temperature $T$ is $v_0=\sqrt{2k_BT/m}$ with $k_B$ the Boltzman constant, $c$ the speed of light and $m$ the atomic mass. $\chi(\nu) = \frac{c}{v_0}\frac{\nu-\nu_0}{\nu_0}$ is a reduced frequency. Finally, to define the Voigt profile, we have introduced the Voigt parameter $a=\Gamma c / (4 \pi v_0 \nu_0)$. The determination of the emission profiles, however, is not straightforward, as they depend on the light initial spectra, but Monte-Carlo simulations show that, in the multiple scattering regime, they are very close to the absorption lines (see Section \ref{Monte-Carlo}).

For Gaussian emission and absorption spectra $k_a(\nu)=k_e(\nu)=k_D(\nu)$, the expected power of the step length distribution asymptotic decay is $\alpha = 2$, while for Lorentzian (Cauchy) spectra $k_a(\nu)=k_e(\nu)=k_L(\nu)$, $\alpha = 1.5$. As Voigt spectra $k_V(\nu)$ show Lorentzian-like frequency wings, they also lead to a decay with $\alpha = 1.5$ \cite{Pereira2004}.

\subsection{Experimental methods and measurements}\label{methods}

\begin{figure}[b]
\begin{center}
\includegraphics[width=8.5cm,height=10.5cm]{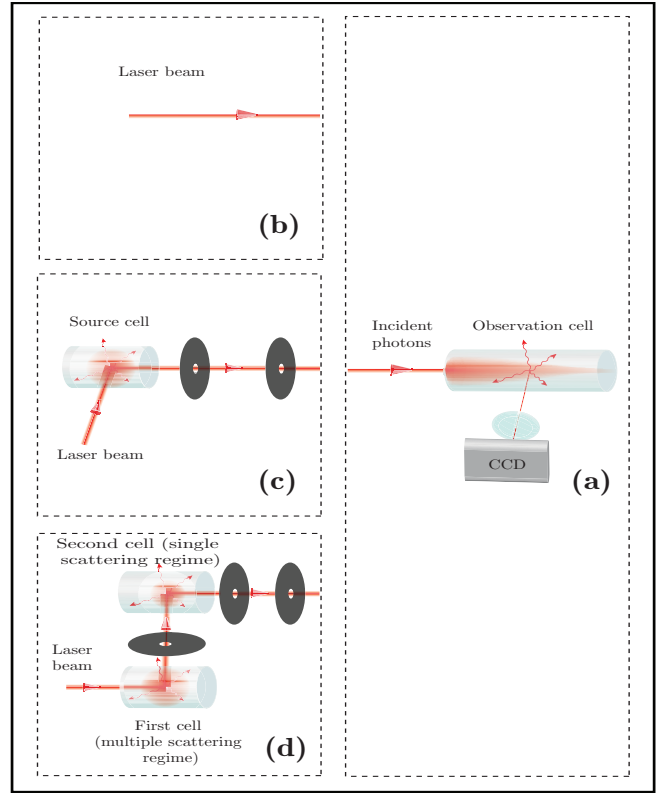}
\caption{Experimental setup for observing the spatial profile of fluorescence in a hot atomic vapor, i.e. the step-size distribution function. (a) Observation cell and imaging system. This part is common to the three performed measurements. (b) The incident photons are directly provided by a monochromatic laser, for calibration. (c) The incident photons have been scattered once, at $90^\circ$, in the low-optical-thickness source-cell. (d) The incident photons have been first scattered several times ($\sim 4$) in a first, optically-thick, cell before being scattered once more in the second cell.}
\label{setup}
\end{center}
\end{figure}

While the measurement of the single step size distribution of a photon in large, hot atomic vapors such as in stars is not feasible, a specific multicell arrangement allowed for such an experimental observation \cite{Mercadier2009}. The experimental arrangement devised to observe individual scattering events in a hot vapor consists of an optical cell filled with a vapor of Rb atoms at a controlled temperature/density, from now on called the observation cell (part (a) of Figure \ref{setup}). These atoms are excited by a collimated beam of photons, so as to define a direction of observation (the ballistic one). The atoms scatter the incident photons in all the directions. Photons detected by a CCD device show the position in the cell where they have been last scattered. A narrow strip (2 mm) parallel to the ballistic direction is selected for measurements. Assuming that photons scattered from this strip to outside the cell have undergone a single scattering event in the observation cell, the measured fluorescence intensity is directly proportional to the step-size distribution $P(x)$. In fact, insuring a single scattering regime in the observation cell is quite incompatible with having a sufficient dynamics of $P(x)$ so that higher order scattering inevitably happens.
As multiple scattering corrections might also mimic a power law intensity along $Ox$, it is crucial to correct for this effect. This can be achieved by subtracting a narrow strip measured slightly off-axis, as those photons are due to multiple scattering only \cite{Mercadier2009}.

A reference measurement is first carried out with a monochromatic laser beam (frequency $\nu_L$) as the incident photon source (See Figure \ref{setup}(b)). The step-size distribution $P(x)$ for this specific configuration is given by Eq. \ref{PdeX} with $\nu = \nu_L$,
\begin{equation}
P(x)\propto e^{-k(\nu_L)x}.
\end{equation}
The measurement of this exponential decay enables the calibration of the mean-free-path $\ell(\nu_L) = 1/k(\nu_L)$ at resonance, and consequently the atomic density $n$, which can be varied through the temperature of the Rb reservoir in the cell. The temperature is varied between 20 and $\approx 47^\circ$C, corresponding to an atomic density between $9 \times 10^{9}$ and $2 \times 10^{11}$ atoms/$\mathrm{cm}^3$ and a subsequent resonant mean free path between 50 and 5 mm.

\begin{figure}[t]
\begin{center}
\includegraphics[width=8.5cm]{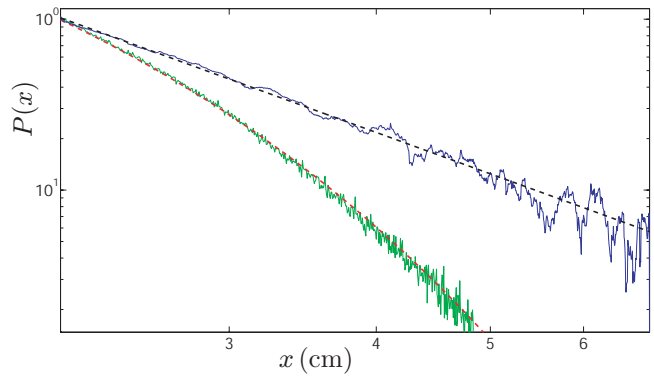}
\caption{Experimental measurements of the step size distribution function (log-log scale) in the configuration of Figure \ref{setup}(a) and (b). With an incident laser beam (green crosses), the observed decay is well fitted by an exponential (red dotted line) as expected from Beer-Lambert law. In the two-cell configuration, the data (continuous blue line) are well fitted by a power law (black dotted line) $P(x)=1/x^{\alpha}$, with $\alpha = 2.41 \pm 0.12$.}
\label{CourbesExp}
\end{center}
\end{figure}

A second experimental measurement of the jump size distribution $P(x)$ is performed by using a two cell configuration (Fig. \ref{setup}(c)). Laser light is sent on a first rubidium cell of very low atomic density, so that photons are scattered at most once in this source cell. Light with a spectra broadened by Doppler effect is emitted in all directions in space, but a collimated beam orthogonal to the initial laser propagation axis is selected out of it by two diaphragms and sent towards the observation cell. The knowledge of the position of the first scattering event, in the source cell, and the second one, in the observation cell, gives us access to the jump size distribution function $P(x)$ (Fig. \ref{CourbesExp}). We find that, for $x$ large enough compared to the mean-free-path of resonant light, $P(x)$ scales like a power law $1/x^{\alpha}$, with $\alpha = 2.41 \pm 0.12$.

Although the obtained distribution has clearly a diverging second moment, this measurement alone does not prove that light transport in the multiple scattering regime is superdiffusive, because the spectrum of light scattered in the first cell, and hence the jump size distribution for the first step, depends on the laser frequency. The frequency redistribution is only partial \cite{Pereira-2008}. To characterize the multiple scattering regime, we need to measure the jump size distribution function after several steps, once photons have lost memory of the initial laser frequency, so that the emission spectra has converged towards a limit spectra (see section \ref{Monte-Carlo} for a numerical demonstration of this convergence). To achieve this, we use a three cell configuration (Fig. \ref{setup}(d)). Laser photons are sent in a first rubidium cell of high atomic density, where they are scattered several times ($\sim 4$), before reaching a second source cell, where they undergo one more scattering event with a well-known position before being sent to the observation cell. We thus measure the jump size distribution function in the multiple scattering regime. Once again, we find a power law asymptotic behavior $P(x) \sim 1/x^\alpha$, with $\alpha = 2.09 \pm 0.15$, which has an infinite variance and is therefore characteristic of a superdiffusive regime. One should notice, however, that as $\alpha>2$ the mean free path remains finite.

\section{Monte Carlo simulations}
\label{Monte-Carlo}

A Monte Carlo (MC) code simulates the spectral and spatial properties of individual photons as they propagate in an atomic vapor at temperature $T$, in an optical cell having the dimensions of the experimental observation cell. Monte Carlo simulations may in general help in the signal deconvolution or in choosing suitable experimental parameters. In this work, it proved indeed useful to validate the data analysis and for a deeper understanding of the third experimental configuration, in the multiple scattering regime (Fig. \ref{setup}(d)).

\subsection{Description}

The code consists essentially in an elementary procedure, corresponding to a scattering order, which is repeated as long as the simulated photon is inside the optical cell. When it gets out, it is either discarded or detected, depending on the direction it takes. If detected, its last scattering position in the cell is recorded in a file, together with the order of the last scattering process. The elementary procedure consists in taking one photon, of frequency $\nu_1$ and wavenumber $\vec k_1$, and to determine the position where it will undergo a scattering process. The distance from this position to the initial one is frequency-dependent. The probability $T(x,\nu)$ that a photon of frequency $\nu$ travel a distance $x$ without being absorbed in the vapor is randomly attributed a value between 0 and 1 and the step length $x$ is then determined as
\begin {equation}
x = -\frac {\ln T}{k(\nu)},
\label{ell}
\end {equation}
with $k(\nu)$ the absorption spectrum. A new wavenumber $\vec k_2$ (direction) is then drawn for the scattered photon and its frequency $\nu_2$ is calculated from
\begin{equation}
\nu_2=\nu_1+\vec v \cdot (\vec k_2 -\vec k_1),
\label{DeltaNu}
\end{equation}
where $\vec v$ is the velocity of the scattering atom. The frequency $\nu_2$ will then determine the step length to the next scattering process.

The photons incident on the optical cell are initially homogeneously distributed over a disk whose diameter is the one of the second diaphragm of the setup shown in Figure 1(c,d). Their direction is parallel to the longitudinal axis of the observation cell and their frequency can be drawn either from a laser's lorentzian lineshape (calibration experiment, see text) or from a Voigt distribution at the temperature of the source cell.

 From the equation (\ref{DeltaNu}), we can see that the relevant components of both the wavenumbers and of the atom velocity are the one parallel and the one in the plane perpendicular to the incident photon's direction.
Regarding the direction of the scattered photon, an isotropic angular distribution follows from isoprobabilistically drawing the value of the polar angle's cosine between $-1$ and $+1$ and the value of the azimuthal angle between $0$ and $2\pi$.

Finally, the velocity component $v_p$ of the scattering atom in the plane perpendicular to the direction of the incident photon can take any value, with a probability law given by the Gaussian Maxwell-Boltzmann distribution of standard deviation $\sigma_v=\sqrt{k_B T/m}$: $G(v_p)\propto e^{-(v_p/\sqrt{2}\sigma_v)^2}$. The velocity component $v_k$ of the scattering atom in the direction of the incident photon is drawn from a probability law $p(v_k)= L(\nu,v_k)\times G(v_k)$, convolution of the atomic Lorentzian lineshape $L(\nu,v_k)\propto 1/(1+4\pi(\nu-\nu_0-v_k/\lambda)^2/\Gamma^2)$ and of the Gaussian Maxwell-Boltzmann velocity distribution $G(v_k)$, where $\nu_0$, $\lambda$ and $\Gamma$ are respectively the frequency, the wavelength and the linewidth of the atomic transition considered.

\begin{figure}[b]
\begin{center}
\includegraphics[width=8.5cm]{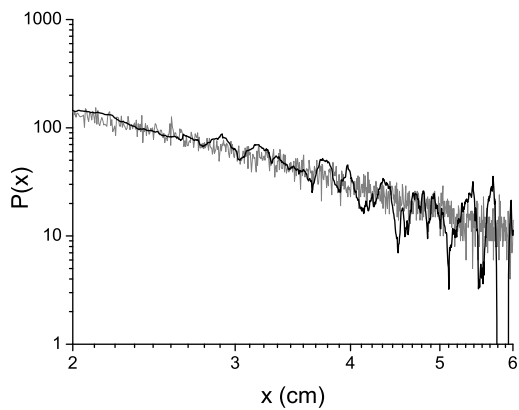}
\caption{First-step size distribution function (log-log scale) in the configuration of Figure \ref{setup}. Black solid line: Experimental measurements, as in Figure \ref{CourbesExp}. Grey line: Monte Carlo simulation.}
\label{MC_Exp}
\end{center}
\end{figure}

The simulated first-step distribution is shown in Figure \ref{MC_Exp} to reproduce the observed one.

\subsection{Multiple scattering regime}
\label{CFR}

\begin{figure}[b]
\includegraphics{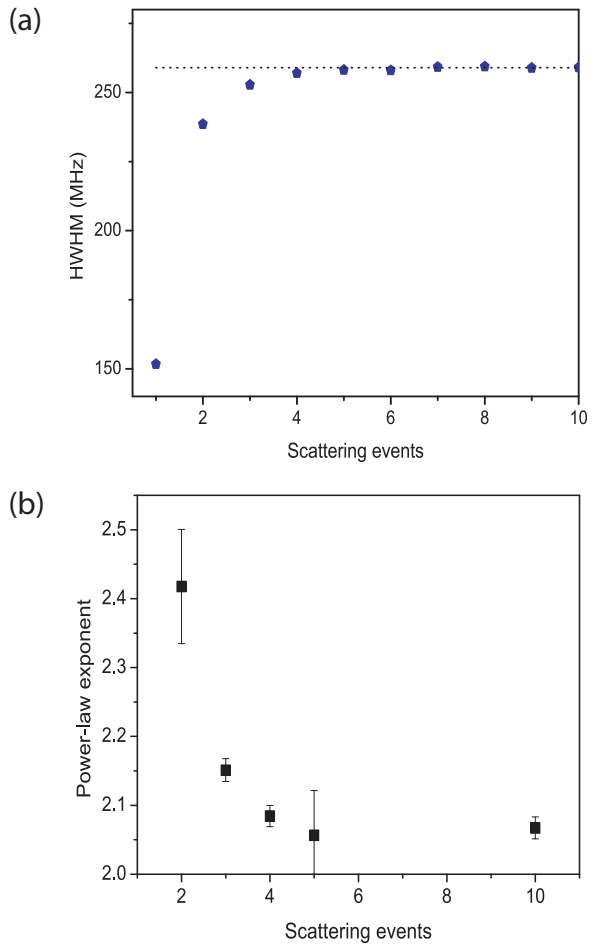}
\caption{Evolution of (a) the HWHM of the photons spectral distribution and (b) the power-law exponent $\alpha$, with the number of scattering events in a Rb vapor at 300K. The original spectral distribution (photons absorbed in the zeroth order diffusion) is the one of a resonant, monochromatic laser.}
\label{Evolution}
\end{figure}

To characterize the multiple scattering regime from the measurement of a single step-size distribution in the middle of the scattering sequence, it is necessary that this distribution have reached a stationnary state and do not depend any more of the photon frequency at the previous step. To know how many scattering events are required to fulfill this condition, we have used the MC simulations to follow the evolution of the light spectral distribution, from the monochromatic laser beam (lorentzian shape of width $\approx$ 1 MHz) to the asymptotical Voigt distribution (Doppler width for the preparation cell temperature). The half width at half maximum (HWHM) of the spectra is shown as a function of the diffusion order in Figure \ref{Evolution}(a). We can see that in the conditions of the experiment (300 K) at least 5 diffusions are necessary for assuring a Voigt-broadened distribution of the photons incident on the observation cell and fulfill the Complete Frequency Redistribution condition. The second setup used in the experiments (see Figure \ref{setup}(d)) takes this requirement into account. The exponent of the power-law of the corresponding step-distribution is shown in Figure \ref{Evolution}(b). It decreases from $\alpha \approx 2.4$ to $\approx 2.05$ as the diffusion order increases from 1 to $\sim 5$.

In our experimental conditions, the Doppler width is typically two orders of magnitude larger than the natural linewidth (Voigt parameter $a\approx 10^{-2}$). In such a regime, the lorentzian nature of the natural broadening is expected to dominate the spectral dynamics (asymptotical Lorentz power-law exponent $\alpha$ = 1.5) for optical densities $n\sigma_0 x$ larger than $10^4$ \cite{Pereira2004}, i.e. for jump sizes larger than 1 m in our experimental conditions ($\sigma_0\approx 1.25\times 10^{-13}$ m$^2$, $n\approx 5 \times 10^{16}$ m$^{-3}$). Below 10 cm, the spatial range observed in our experiment, the spectral behavior (assuming CFR) is purely Doppler-driven, with an exponent $\alpha = 2$ for the step distribution power law.

\section{Conclusion}
We have evidenced a L\'{evy} Flights behavior of photons propagation in a hot resonant atomic vapor by measuring the single step distribution in this multiple scattering process. Monte Carlo simulations help deconvoluating the effects of multiple scattering and analyzing the frequency redistribution process in a Doppler-broadened vapor excited by a monochromatic light source.

\end{document}